# Uranium at High Pressure from First Principles

S. Adak[a,b], H. Nakotte[a,b], P.F. de Châtel[a], B. Kiefer[a]

[a]*Department of Physics, New Mexico State University, Las Cruces, NM 88003, USA*

[b]*Los Alamos Neutron Science Center, Los Alamos National Laboratory, Los Alamos, NM 87545, USA*

**Abstract**

The equation of state, structural behavior and phase stability of α-uranium have been investigated up to 1.3 TPa using density functional theory, adopting a simple description of electronic structure that neglects the spin-orbit coupling and strong electronic correlations. The comparison of the enthalpies of *Cmcm* ($\alpha$-U), *bcc*, *hcp*, *fcc*, and *bct* predicts that the $\alpha$-U phase is stable up to a pressure of ~285 GPa, above which it transforms to a *bct*-U phase. The enthalpy differences between the *bct* and *bcc* phase decrease with pressure, but *bcc* is energetically unfavorable at least up to 1.3 TPa, the upper pressure limit of this study. The enthalpies of the close-packed *hcp* and *fcc* phases are 0.7 eV and 1.0 eV higher than that of the stable *bct*-U phase at a pressure of 1.3 TPa, supporting the wide stability field of the *bcc* phase. The equation of state, the lattice parameters and the anisotropic compression parameters are in good agreement with experiment up 100 GPa and previous theory. The elastic constants at the equilibrium volume of $\alpha$-U confirm our bulk modulus. This suggests that our simplified description of electronic structure of uranium captures the relevant physics and may be used to describe bonding and other light actinides that show itinerant electronic behavior especially at high pressure.

**Corresponding author:**

Boris Kiefer

Physics Department, MSC3D

New Mexico State University

Las Cruces, NM 88003

Phone: 575-646-1932; Fax: 575-646-1934

E-mail: bkiefer@nmsu.edu

# 1. Introduction

Uranium (U) is an element of great importance as it occurs in many diverse environments. For example, it is one of the main nuclear fuel materials [1] and it has been speculated also to be a source of the internal heating of our planet [2]. An understanding of the electronic structure of uranium can provide models for uranium in different bonding environments. The *5f* electrons in the actinide metals often exhibit interesting behavior under pressure. For example, the *5f* electrons in the elements Th–Pu (including uranium) are itinerant and contribute to the bonding [3]. In contrast, a more localized and non-bonding behavior of the *5f* states, similar to the *4f* states lanthanide series, is found throughout the remainder of the elements in the actinide series. The study of the electronic structure of light actinides under pressure yields important information about the role of the *5f* electrons in these materials, and it may provide fundamental insight into electronic structure and phase stability under high compression.

Uranium has a very rich phase diagram. A previous study on the phase diagram of uranium by Yoo *et al.* [4] reports that uranium crystallizes in the orthorhombic structure (α - phase) with space group *Cmcm* below 935 K temperature above which, at ambient pressure, the phase of uranium changes to the *β*-phase (*bct*), which is transforms to the *γ*-phase (*bcc*) above 1045 K. The *β-γ* phase transition has a very low Clapeyron slope and the stability field of *β*-U terminates in a triple point at ~3 GPa and ~1000 K [4]. At lower temperatures, *α*-U was reported to be stable up to at least 70 GPa [4,5] while the *γ*-U is found to be stable for higher temperatures. The phase boundary between *α*-U and *γ*-U flattens with increasing pressure and the stability field of *γ*-U widens, which was attributed to the comparatively low bulk modulus of the *bcc* phase [4]. Additional pressure studies by Akella *et al.* [6] and Le Bihan *et al.* [7] led to the conclusion that *α*-U is the most stable phase up to at least 100 GPa at room temperature.

In general, heavy elements pose a large challenge to electronic structure theory due to relativistic effects because of the large nuclear charge and the often complex hybridization of *s*-, *p*-, *d*- and *f*-electrons that are difficult to treat self-consistently. On the other hand, previous pseudopotential studies by Richard *et al.* [8] and all-electron computations by Söderlind [9] revealed that the contributions due to spin-orbit coupling are relatively small for the light actinides, up to and including uranium. Hence, in the case of uranium, a simpler description of electronic structure may suffice and is worth testing.

This paper presents the results of calculated structural behavior and equation of state (EOS) of uranium up to 1.3 TPa. We also present a phase stability study on α-U against four other phases namely *bcc*-U, *fcc*-U, *hcp*-U and *bct*-U. Our studies go beyond previously published work in that a) we optimize all lattice and internal degrees of freedom simultaneously in order to capture the correct energetics in uranium; b) we investigate the enthalpies of all considered phases in order to gain further insight into the ultimate high-pressure phase of uranium; c) we provide an explanation of the poorly understood anisotropic compression of α-U; and d) we provide an explanation of the unusual large range of EOS parameters observed in experiments. However, we did compare our findings with published (experimental or theoretical) work at lower pressures, whenever available. As will be shown below, we find that a simplified electronic structural model works well and captures the relevant physics for uranium, including phase stability, equation of state and anisotropic compression.

## 2. Computational method

All the first-principles calculations were executed within the framework of density functional theory (DFT) of Hohenberg and Kohn [10] to study the EOS and phase stability of Uranium. In this study, we investigated five potential phases of uranium: (a) the orthorhombic *α-U* with space group *Cmcm*, (b) the *bcc*-U with space group $Im\bar{3}m$, (c) a hypothetical cubic *fcc*-U with space group $Fm\bar{3}m$, (d) a hypothetical *hcp*-U with space group $P6_3/mmc$ and (e) the *bct*-U with space group *P4/mmm* (Fig. 1). The one-electron wave functions were calculated by means of the projector augmented wave (PAW) formalism [11,12]. The exchange and correlation potential was used as parameterized in Perdew *et al.* [13]. The core radius used for Uranium was 1.811 Bohr (valence configuration: [Xe core]$5f^3 6s^2 6p^6 6d^1 7s^2$). All calculations were performed using the Vienna *ab-initio* simulation package (VASP) [14]. Convergence tests showed that a plane-wave cut-off energy of 600 eV and standard *k*-point grids [15] of 8x8x8 for *α*-U, 12x12x10 for *hcp*-U and 14x14x14 for *fcc*-U, *bcc*-U and *bct*-U were sufficient to achieve convergence in total energies and pressures to within 3 meV/atom and 0.1 GPa, respectively. These *k*-point grids are similar to those used in previous plane-wave studies [8]. To determine the hydrostatic ground states of the studied phases, geometry optimizations were performed by relaxing lattice parameters and internal coordinates at fixed volume. The equilibrium geometries were determined by relaxing each phase at constant pressure using the same computational parameters. We adopted the simplest electronic structure model which neglected spin-orbit coupling and strong electronic correlations (Hubbard-U). All-electron methods that have previously been used to investigate elemental uranium [7-9,16-18] adopt a split basis-set approach, a plane-wave basis set in the interstitial region between atoms and localized basis functions within the muffin-tin spheres. The size of the basis-sets is chosen such that the spheres overlap as little as possible. However, the spatial shape of the *f*-orbitals can cause a problem with this splitting since they may "peak" through the Muffin-tin sphere into the interstitial space and hence it may be difficult to treat *f* electrons in a split basis set methodology. In contrast, we used a plane-wave basis-set which works well for extended orbitals. This approach benefits from the itinerant character of the *5f* electronic states in the light actinides, including uranium [1]. All of our calculations were static (performed at 0 K) and neither zero-point motion nor vibrational contributions were included. The EOS and phase stability of uranium were studied in the pressure range 0 – 1.3 TPa. The EOS parameters – volume ($V_0$), bulk modulus ($B_0$), and the first pressure derivative of the bulk modulus $\left(B_0^{'}\right)$ – were obtained from a third-order Birch-Murnaghan EOS [19]:

$$E = E_0 + \frac{9B_0V_0}{2}\left\{\frac{1}{2}\left[\left(\frac{V}{V_0}\right)^{-\frac{2}{3}} - 1\right]\right\}^2 \times \left[1 + \left(B_0^{'} - 4\right)\left\{\frac{1}{2}\left[\left(\frac{V}{V_0}\right)^{-\frac{2}{3}} - 1\right]\right\}\right], \quad (1)$$

where $E_0$ is the equilibrium energy. Elastic constants were computed by applying small strains of magnitude ±1% to the relaxed structure. Subsequently, the lattice was kept fixed and the internal coordinates were re-relaxed. Using a linear stress-strain relationship, we obtained the complete elastic constant tensor from the stress response to the applied strain.

### 3. Results and discussions

Figure 2 shows the volume compression for the α-U phase. For comparison, data points from earlier theoretical and experimental studies (Le Bihan *et al.* [7], Akella *et al.* [6], Yoo *et al.* [4], and Virad *et al.* [20]) are also shown. To extract the EOS parameters, the calculated pressure-volume (PV) data points were fitted using the third-order Birch-Murnaghan (BM) equation of state [19]. In Table 1, we list the values for the EOS parameters as obtained by fitting the energy-volume data, together with previous experimental and theoretical results [4-9,16,21-23]. As can be seen from Table 1 and Figure 2, the agreement of our results with experiments is excellent below 50 GPa, but they deviate slightly from the experimental data points reported by Akella *et al.* [6] and Yoo *et al.* [4] for higher pressures.

Our value for the unit-cell volume, $V_0$ = 20.19 Å$^3$/atom, is slightly lower than the experimental values [7,21,22], which is expected since we have neglected the thermal expansion in our theory. Using a thermal volume expansion coefficient for uranium of 39.0 x 10$^{-6}$ K$^{-1}$ [4], we predict that this temperature difference increases the volume to 20.43 Å$^3$ at room temperature, in somewhat better agreement with experiments.

Our calculated value for the bulk modulus, as obtained from the E(V) fit for the *Cmcm* phase, is 135.5 GPa, in very good agreement with previous all-electron calculations with or without spin-orbit coupling [8,9]. However, our bulk modulus is ~30% higher compared to the experimental value (104 GPa) reported by Le Bihan *et al.* [7]. On the other hand, we note that Le Bihan's value for $B_0$ (Table 1) is significantly lower than other experimental values, and our value is well within the range of reported values (104 – 147 GPa). The value for the first derivative of the bulk modulus as obtained from our calculations ($B_0^{'}$ = 4.97) is also within the range (3.78 – 6.20) of previously reported values. The well-known anti-correlation of $B_0$ and $B_0^{'}$ greatly improves the agreement with Le Bihan's experimental value of the bulk modulus, i.e. fitting our EOS using Le Bihan's value for $B_0^{'}$ = 6.2 [7] leads to a bulk modulus of 110.9 GPa, in good agreement with Le Bihan's low experimental value. Thus, the perceived low bulk modulus of the *α*-U phase can largely be attributed to the trade-off between $B_0$ and $B_0^{'}$. This is further corroborated by considering the EOS parameters ($E_0$, $V_0$, $B_0$ and $B_0^{'}$) for all five phases The values are tabulated in Table 2, where we also included the theoretical values reported by Richard *et al.* (with and without spin-orbit coupling) [7] and Söderlind *et al.* [9] as well as experimental data from Yoo *et al.* [4] and Barrett *et al.* [21] for comparison. In the table, we list two sets our EOS parameters: the first column lists the values determined at pressures up to 100 GPa and the second column contains the values at pressures up to 1.3 TPa. This was done in order to allow a better comparison with previous data [4,8,9,21], which were limited to pressures up to 100 GPa. In the 1.3 TPa EOS parameter column of Table 2, we also included their values as obtained with $B_0^{'}$ fixed to their low-pressure fit values, and those refitted EOS parameters are given in brackets. We find that our EOS parameters are in good agreement with previously published data. We find that the BM-EOS for the low-P fit agree reasonably well with the values derived when the pressure range is increased 10 times. Thus, a BM-EOS appears to be a suitable extrapolation scheme even at extreme compression. The largest deviation occurs for the *fcc* phase (Table 2), and its unusual sensitivity to an increased pressure range remains elusive. We note that the bond distances in *fcc* are neither shortest nor longest in comparison to other phases. This can be taken as evidence that changes in hybridization cannot account for these changes. Furthermore, for a fixed volume, the *fcc* phase has no degree of freedom, which eliminates the possibility of multiple structural minima in the potential energy surface. The comparison of the EOS values in Table 2 also shows that possible differences in the parameters can be attributed to trade-offs between fitting parameters, i.e. small changes in $B_0^{'}$ can have a large effect on $B_0$. This further supports that the anomalously low $B_0$ value in Le Bihan *et al.* [7] may be due to the fitting procedure used. A comparison between different phases corroborates that the equilibrium volume, $V_0$, is smallest for the *Cmcm* phase, while its 100-GPa bulk modulus, $B_0$, is higher compared to the considered *bcc*, *fcc*, *hcp* and *bct* phases of uranium.

In Table 3, we have listed our computed elastic constant tensor, which seem to show better agreement with the experimental values by Fisher and McSkimin [24] compared to previous theoretical study by Söderlind *et al.* [9]. Using $\sqrt{\frac{1}{9}\sum_{i=1}^{9}\left(cij_{Theory} - cij_{Experiment}\right)^2}$ as a metric, our average deviation from experimental values of 22 GPa is less than the 39 GPa for Söderlind *et al.* [9], who used coupled strains in their study. The re-computed Voigt-Reuss-Hill averaged bulk modulus is in excellent agreement with the values obtained from the EOS fits, which further validates our simplified approach to describe the electronic structure of uranium.

The variation of the axial ratios as a function of reduced volume for the *α*-U phase as obtained from our calculations is shown in Figure 3. For comparison, experimental and theoretical data points from a previous study (Le Bihan *et al.* [7]) have also been plotted. It is evident that our calculated *b/a, c/a* and *b/c* axial ratios are in good agreement with the previous experimental and theoretical values.

Figure 4 shows the calculated absolute lattice parameters for the *α*-U phase as a function of reduced volume, together with the data points of Le Bihan *et al.* [7]. The figure indicates a very good agreement with the previously reported values, except that the variation of the *c*-axis shows deviations from the experimental values but agrees with the previous computations in the same study. The relative changes in the lattice parameters as function of reduced volume is shown in Figure 5. Our results indicate that up to 400 GPa, the *c*-axis is least compressible, consistent with previous experiments and theory [7]. In the *α*-U structure (see Figure 1), two types of U-U bonding occur, one sub-parallel to the *c*-axis and the other along the *a*-axis. The reported bond lengths from experiments along *c*-axis direction (2.794 Å) are shorter than the bond lengths (2.854 Å) in *a*-axis direction [25]. Since the linear length scales as $\propto \sqrt[3]{V}$, shorter bonds may be expected to have a lower compressibility. The experimental studies by Le Bihan *et al.* [7] also revealed that the *a*-axis is softer than the *b*-axis at fairly low pressure (Figure 5). Simple cube fits of the lattice parameters/pressure data to the BM-EOS (Eq. 1) result in linear moduli of 642 GPa, 287 GPa and 350 GPa for the *c*-, *a*- and *b*-axis, respectively. The similar *a*- and *b*-axis moduli may raise the expectation that the two axes behave similarly. However, pressure derivative of the *a*-axis is almost twice that of the *b*-axis. As a result the *a*-axis predicted to be softest below ~25 GPa, where it crosses the compression curve of the *b*-axis and its behavior is similar to the incompressible *c*-axis for higher pressures in excellent agreement with the FPLMTO computations in Le Bihan *et al.* [7]. The re-calculated volumetric bulk modulus from the linear moduli (127 GPa) is in reasonable agreement with $B_0$ (135.5 GPa, Table 1).

To study the phase stability of the four other potential phases, we computed the difference (relative to the orthorhombic *α*–U phase) in enthalpy, ΔH, for all the phases at different pressures up to 1.3 TPa. Figure 6 shows the variation of the difference in enthalpy with respect to the *α*-U phase for *bcc, fcc, hcp* and *bct* phases, as function of pressure. In the analysis, we omitted the computed *α*-U point at 1290 GPa, where the analysis of the relaxed cellshape and the internal coordinates show that the *α*-U phase has relaxed toward the *bct*-U phase. In contrast, at 860 GPa the same analysis shows that *a*-U remains orthorhombic. As can be seen, all enthalpy differences with respect to *α*-U are positive in the pressure range below ~285 GPa, in agreement with previous experiment up to 100 GPa [4,5] and theory [8,9,16,17]. At this pressure, our results predict a first-order phase transition from *α*-U ➔ *bct*-U with a volume decrease of ~0.9%. We note that our transition pressure of ~285 GPa is lower than the previously predicted transition pressure of 363 GPa [26], which may be due to the fact that we consider relaxation of the axial ratios for the *α*–U and *bct*-U phases in our study, while they were neglected in Li and Wang's study [26]. The occurrence of such a structural transition has been discussed previously by Richard *et al.* [8] and Söderlind *et al.* [27]. With increasing pressure, we find that enthalpy difference between *bcc*-U and *bct*-U decreases at a slow rate of $0.933 \times 10^{-3}$ eV/100 GPa at least to 1.25 TPa, in agreement with Li and Wang [26]. Linear extrapolation of the enthalpy differences between 900 GPa and 1200 GPa gives a *bct* ➔ *bcc* transition pressure of ~2.3 TPa. On the other hand, in an isochoric ensemble, we predict that the phase transition occurs at a compression $V/V_0 = 0.58$, in reasonable agreement with the previously reported range for $V/V_0$ from 0.60 to 0.69 [8,17,18]. The slight variation in $V/V_0$ values originate (at least partly) from the fact that some structural parameters had to be fixed during the geometry optimization in above studies. At least up to 1.3 TPa, the enthalpies of the close-packed *fcc* or *hcp* are higher than the one of *bct*, by 1.0 eV/atom and 0.8 eV, respectively. This indicates that the *bcc*-U phase is remarkably stable and may indeed be the ultimate high-pressure phase of uranium [17,27]. Our results show that the difference in enthalpy for the *bcc*-U phase is smaller than the hypothetical close-packed *fcc*-U and *hcp*-U phase by a significant amount (for comparison: $\frac{3}{2}kT$ amounts to only ~0.04 eV at room temperature).

## 4. Summary

Using the simplest electronic structure model, neglecting spin-orbit coupling and strong electronic correlations, we find that the *α*-U is the stable phase up to at least 285 GPa, above which the *bct* phase becomes stable. We also estimated a transition pressure of ~2.3 TPa for the stabilization of *bcc*-U. The enthalpy differences of the close packed structures remain 0.7 – 1.0 eV at P=1.3 TPa, which indicates the stability of the *bcc*-U phase over a wide pressure range. The equation of state and the lattice parameters, as calculated in this study, are in good agreement with experimental data. Furthermore, our results show that the bulk modulus and its pressure derivative are strongly correlated, which can account for some if not most of the experimentally observed variations in $B_0$ for *α*-U. The re-computed bulk moduli from the elastic constants are in excellent agreement with the values obtained from the EOS fits, which further validates that our simple electronic structure model captures the main features for uranium to high pressures.

**Acknowledgements**

This work was supported by National Science Foundation under Grant No. DMR 0804032. The authors would like to acknowledge insightful discussions with Per Söderlind (LLNL).

**Tables and captions**

**Table 1**.
Comparison of the EOS parameters for *α*-U obtained different experimental and theoretical studies in chronological order. Unless otherwise noted, experimental studies reflect values obtained at room temperature, while theoretical studies are zero-temperature calculations.

| $V_0$ [Å$^3$/atom] | $B_0$ [GPa] | $B_0'$ | Reference |
|---|---|---|---|
| 20.52 | | | Barrett *et al.* (1963) [21] experiment[a] |
| 20.75 | | | Donohue (1974) [22] experiment |
| | 125 | 6.2 | Akella *et al.* (1985) [5] experiment |
| | 147 | 2.8 | Dabos *et al.* (1987) [23] experiment |
| | 138.7 | 3.78 | Akella *et al.* (1990) [6] experiment |
| 19.49 | 172.2 | | Söderlind *et al.* (1994) [16] theory |
| | 135.5 | 3.79 | Yoo *et al.* (1998) [4] experiment |
| 19.91 | 145 | | Richard *et al.* (2002) [8] theory |
| 20.67 | 133 | 5.4 | Söderlind (2002) [9] theory |
| 20.77 | 104 | 6.2 | Le Bihan *et al.* (2003) [7] experiment |
| 20.19 | 135.5 | 4.97 | this study theory |

[a]Experiment was done at 40 K

**Table 2.**
Phase stability of various potential uranium structures, i.e. *α*-U, *bcc*-U, *fcc*-U, *hcp*-U and *bct*-U. Comparison of the EOS parameters, as obtained from BM-fits to the energy-volume data for all studied phases, with previously published data.

| Phase | This study EOS: $P < 100$ GPa | This study EOS: $P < 1.3$ TPa | Richard et al. [8] | Söderlind [9], Söderlind et al. [16] | Experiment |
|---|---|---|---|---|---|
| **α-U** | | | *α-U* [a] | *α-U* [a] | |
| $E_0$ [eV] | -11.525 | -11.525 (-11.52) [b] | | | |
| $V_0$ [Å$^3$]/atom | 20.19 | 20.15 (20.00) [b] | 19.91 (20.33) [c] | 19.49 [d], 20.6 [e] | 20.52 [f] |
| $B_0$ [GPa] | 135.5 | 136.1 (145.7) [b] | 145.0 (132.0) [c] | 172.2 [d], 133.0 [e] | 135.5 [g] |
| $B_0^{'}$ | 4.97 | 5.10 (4.97) [b] | | 5.4 [e] | 3.79 [g] |
| **bcc-U** | | | | | |
| $E_0$ [eV] | -11.320 | -11.313 (-11.330) [b] | | | |
| $V_0$ [Å$^3$]/atom | 20.50 | 20.09 (20.63) [b] | | | 24.33 [g] |
| $B_0$ [GPa] | 116.1 | 140.3 (113.6) [b] | | | 113.3 [g] |
| $B_0^{'}$ | 5.24 | 4.90 (5.24) [b] | | | 3.37 [g] |
| **fcc-U** | | | *fcc*-U | *fcc*-U | |
| $E_0$ [eV] | -11.216 | -11.237 (-11.164) [b] | | | |
| $V_0$ [Å$^3$]/atom | 21.71 | 22.32 (20.60) [b] | 21.3 (21.98) [c] | 20.54 [d] | |
| $B_0$ [GPa] | 115.0 | 85.71 (153.8) [b] | 115.0 (108.0) [c] | 147.9 [d] | |
| $B_0^{'}$ | 4.42 | 5.19 [b] (4.42) | | | |
| **hcp-U** | | | | | |
| $E_0$ [eV] | -11.341 | -11.332 (-11.356) [b] | | | |
| $V_0$ [Å$^3$]/atom | 21.35 | 21.03 (21.57) [b] | | | |
| $B_0$ [GPa] | 105.9 | 121.7 (101.0) [b] | | | |
| $B_0^{'}$ | 5.11 | 4.85 (5.11) [b] | | | |
| **bct-U** | | | | | |
| $E_0$ [eV] | -11.427 | -11.423 (-11.428) [b] | | | |
| $V_0$ [Å$^3$]/atom | 20.61 | 20.34 (20.55) [b] | | | |
| $B_0$ [GPa] | 112.6 | 126.3 (116.2) [b] | | | |
| $B_0^{'}$ | 5.22 | 5.08 (5.22) [b] | | | |

[a] fixed internal coordinates (4*c*-site; space group 63, *Cmcm*); [b] refitted EOS with $B_0^{'}$ fixed at its low-pressure value; [c] including spin-orbit coupling; [d] ref. 9; [e] ref. 16; [f] ref. 21; [g] ref. 4,

**Table 3.**
Elastic constants of $\alpha$-U at the equilibrium volume. The Voigt-Reuss-Hill averaged bulk modulus is also listed. All quantities are in GPa.

| Elastic parameter | Experiment Fisher & McSkimin [24] | Theory Söderlind [9] | Theory Our study |
|---|---|---|---|
| $C_{11}$ | 215.0 (210.0)[a] | 300.0 | 230.1 |
| $C_{12}$ | 46.5 | 50.0 | 69.8 |
| $C_{13}$ | 21.8 | 5.0 | 37.6 |
| $C_{22}$ | 199.0 (215.0)[a] | 220.0 | 196.6 |
| $C_{23}$ | 108.0 | 110.0 | 134.3 |
| $C_{33}$ | 267.0 (297.0)[a] | 320.0 | 312.4 |
| $C_{44}$ | 124.0 (145.0)[a] | 150.0 | 130.4 |
| $C_{55}$ | 73.4 (94.5)[a] | 93.0 | 93.4 |
| $C_{66}$ | 74.3 (87.1)[a] | 120.0 | 82.3 |
| $B_{VRH}$ | 113.2 | 129.3 | 133.2 |

[a]Experimental elastic parameter, extrapolated to T = 0 K (after Söderlind [9])

**Figure captions**

**Figure 1.**
Uranium structures. The volume is 20.8 Å$^3$/atom. Solid lines show U-U bonds that are shorter than 3.1 Å. a) *α*-U, b) *bcc*-U, c) *fcc*-U, d) *hcp*-U and e) *bct*-U. In the case of *α*-U, the two shortest U-U bonds are shown. The shortest bond is sub-parallel to the *c*-axis (vertical) and the second bond is parallel to the a-direction; the next shortest bond is ~15% longer. In the case of *bct*-U, the shortest bond is parallel to the *c*-axis (vertical), the second bond extends along the body-diagonal of the unit cell and is ~19% shorter than the next bond. Other bond distances for *bcc*, *fcc* and *hcp* can be calculated using conventional textbook knowledge.

**Figure 2.**
Volume compression of *α*-U. This study: solid circles and solid line; Söderlind [9], theory: dashed line; Le Bihan *et al.*, theory [7]: solid left triangles; Akella *et al.*, experiment [5]: open circles; Yoo *et al.*, experiment [4]: open squares; Viard *et al.*, experiment [20]: open downward triangles: Viard et al. (1962); Le Bihan *et al.* [7], experiment (N$_2$ pressure medium); upward triangles; Le Bihan *et al.* [7], experiment (Si oil): right triangles.

**Figure 3.**
Axial ratios as a function of volume compression. Dashed lines are guides to the eye. This study: solid circles, solid squares, and solid diamonds for *b/a*, *c/a*, and *b/c*, respectively; Le Bihan *et al.* [7], experiment: open upward triangles, right triangles and downward triangles for *b/a*, *c/a*, and *b/c*, respectively. Le Bihan *et al.* [7], theory: corresponding solid symbols.

**Figure 4.**
Lattice parameters of *α*-U as a function of volume compression. This study: solid circles and dashed lines; Le Bihan *et al.* [7], theory: solid left triangle; Söderlind [9], theory: solid squares); Le Bihan *et al.* [7], experiment (N$_2$ pressure medium): open upward triangles; Barrett *et al.* [21], experiment: crossed circles.

**Figure 5.**
Compression of lattice parameters of *α*-U. This study: solid circles, *a*-axis; solid squares, *b*-axis; solid diamonds, *c*-axis. Dashed lines, guides to the eye, this study. Le Bihan *et al.* [7], experiment: upward triangles, *a*-axis; downward triangles, *b*-axis; right triangles, *c*-axis; Le Bihan *et al.* [7], theory: corresponding closed symbols. Inset: this study, pressure dependence of lattice parameters.

**Figure 6.**
Enthalpy difference of *bct* (solid), *bcc* (dotted), *hcp* (short dashed), and *fcc* (long dashed) uranium phases relative to *α*-U.

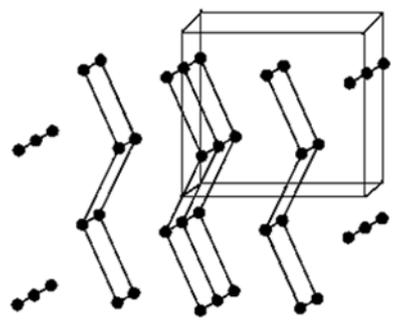

a) α-U; Space group: *Cmcm*

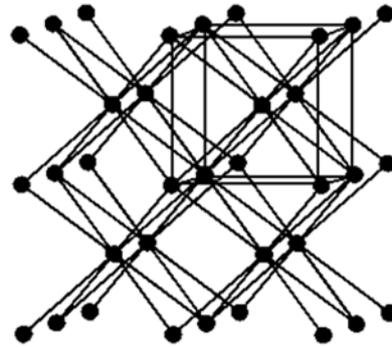

b) *bcc*-U; Space group: $Im\bar{3}m$

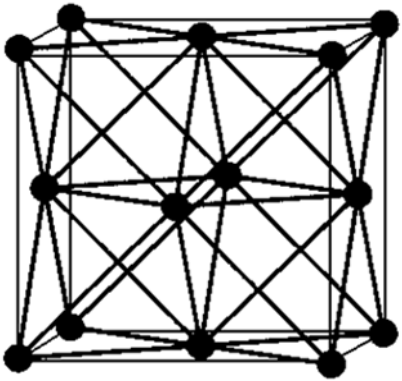

c) *fcc*-U; Space group: $Fm\bar{3}m$

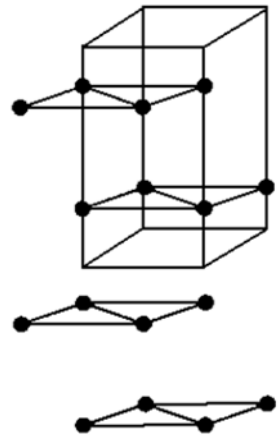

d) *hcp*-U; Space group: $P6_3/mmc$

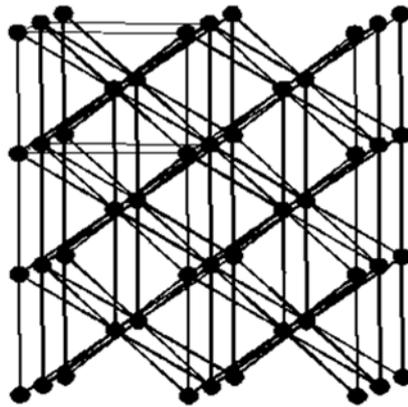

e) *bct*-U; Space group: $P4/mmm$

**Figure 1**

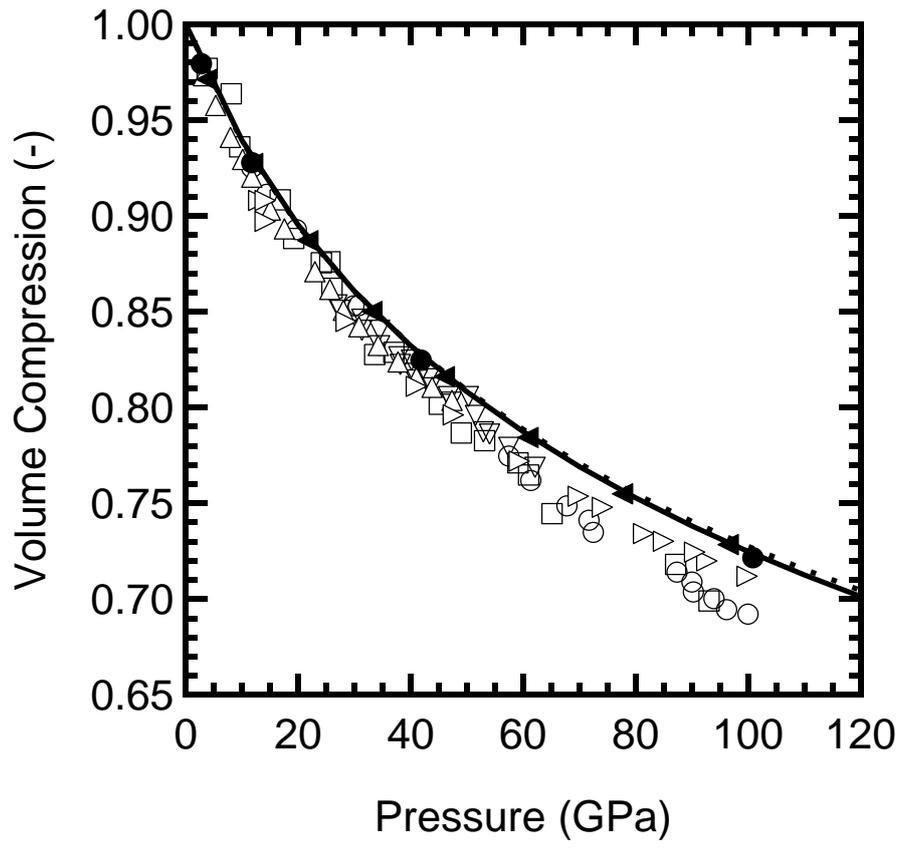

**Figure 2**

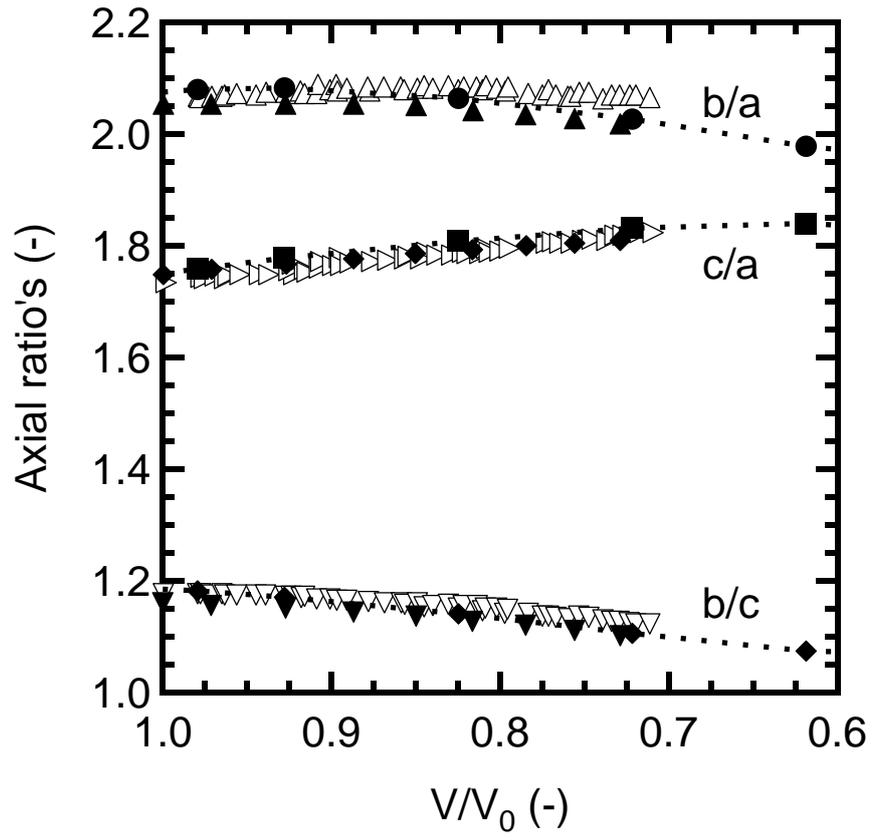

**Figure 3**

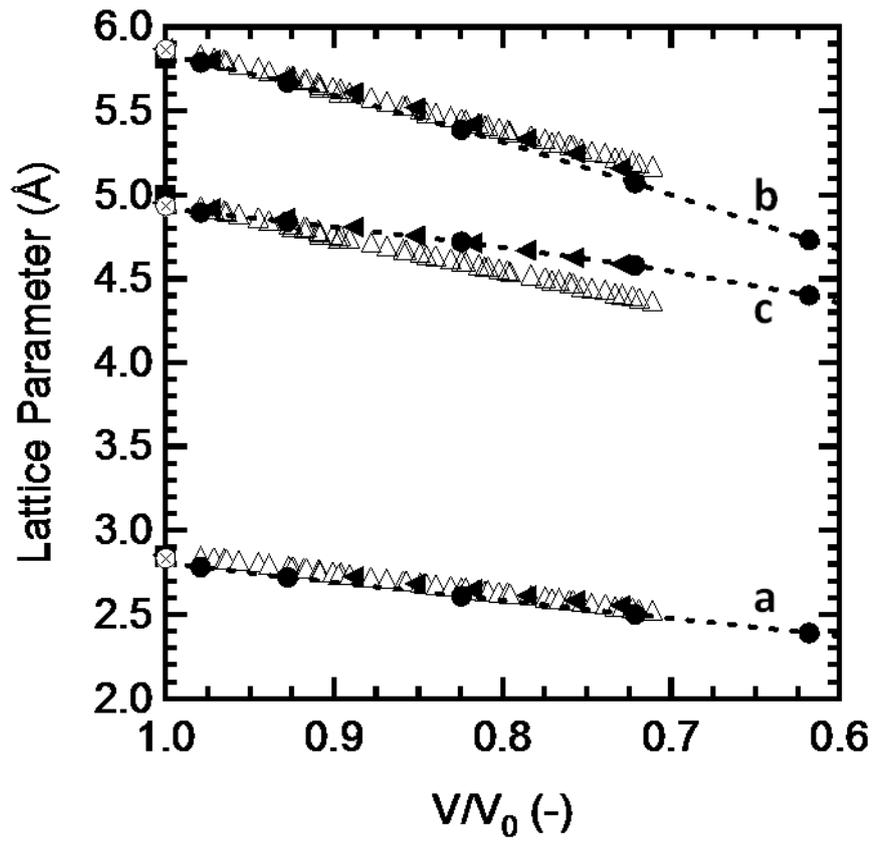

**Figure 4**

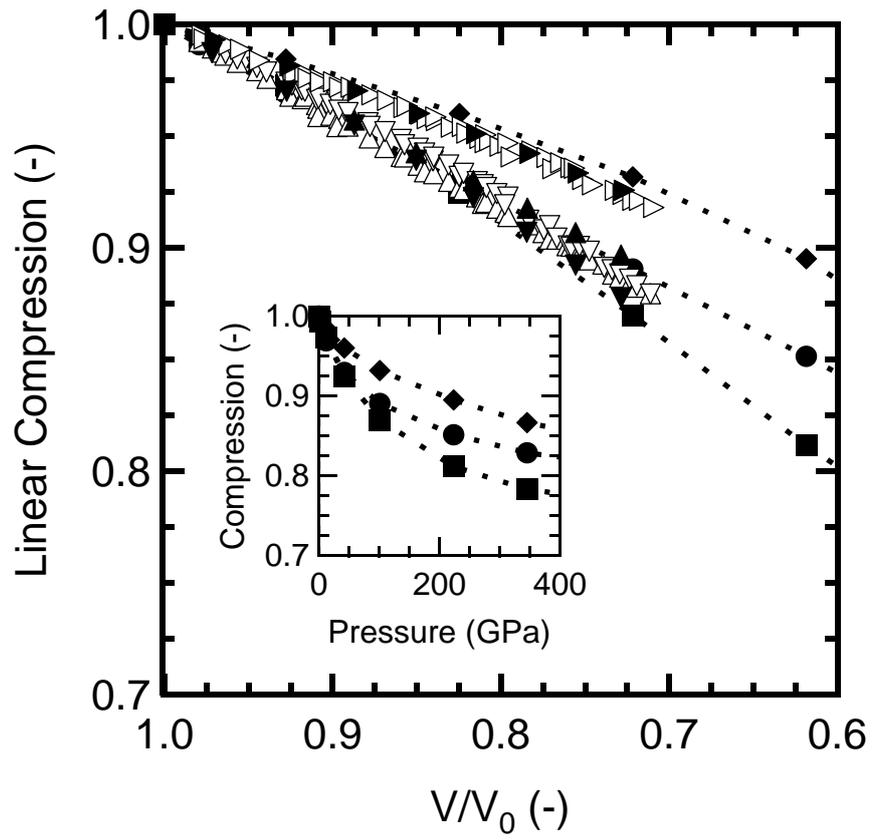

**Figure 5**

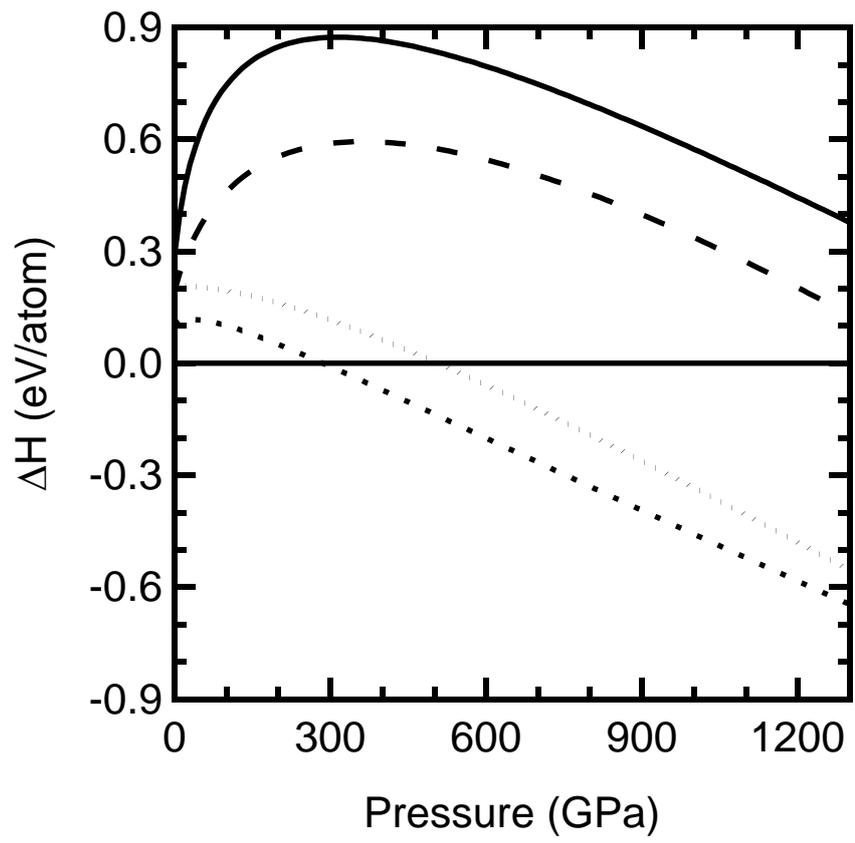

**Figure 6**